\documentclass[aps, nofootinbib,floatfix, amsmath, amssymb, runinaddress, tightenlines, 11pt]{revtex4-1}
\pdfoutput=1
\usepackage[utf8]{inputenc}
\usepackage{amsmath}
\usepackage{xcolor}
\usepackage{colortbl}
\usepackage{multirow}
\usepackage{amssymb}
\usepackage{graphicx}
\usepackage[margin=1in]{geometry}
\usepackage{subcaption}
\usepackage{wrapfig}

\usepackage{amssymb,amsmath,amsfonts,amsbsy,epsfig,wrapfig,caption,float,slashed,graphicx, pstricks}
\usepackage[singlelinecheck=off,justification=raggedright]{caption}

\def\simpropto{\lower.2ex\hbox{$\; \buildrel \propto \over \sim \;$}}
\def\ltsim{\lower.5ex\hbox{$\; \buildrel < \over \sim \;$}}
\def\gtsim{\lower.5ex\hbox{$\; \buildrel > \over \sim \;$}}

\begin{document}
\author{Philippa S. Cole$^{1}$}\email{p.cole@sussex.ac.uk}
\author{Joseph Silk$^{2,3,4}$}\email{silk@iap.fr}
\affiliation{\\ 1) Department of Physics and Astronomy, University of Sussex, Brighton BN1 9QH, United Kingdom}
\affiliation{\\ 2) Institut d'Astrophysique de Paris, UMR 7095 CNRS, Sorbonne University, 75014 Paris, France}
\affiliation{\\ 3) Department of Physics and Astronomy, Johns Hopkins University,
Baltimore, MD 21218, USA}
\affiliation{\\ 4) Beecroft Institute of Particle Astrophysics and Cosmology, Department of Physics, University of Oxford, Oxford OX1 3RH, UK}
\date{\today}
\title{Small-scale primordial fluctuations in the 21cm Dark Ages signal}

\begin{abstract}
Primordial black hole production in the mass range $10-10^4 \,{\rm M_\odot}$ is motivated respectively by interpretations of the LIGO/Virgo observations of binary black hole mergers and by their ability to seed intermediate black holes which would account for the presence of supermassive black holes at very high redshift. Their existence would imply a boost in the primordial power spectrum if they were produced by overdensities reentering the horizon and collapsing after single-field inflation. This, together with their associated Poisson fluctuations would cause a boost in the matter power spectrum on small scales. 
The extra power could become potentially observable in the 21cm power spectrum on scales around $k\sim0.1 - 50\,{\rm Mpc^{-1}}$ 
with the new generation of filled low frequency interferometers. We explicitly include the contribution from primordial fluctuations in our prediction of the 21cm signal which has been previously neglected, by constructing primordial power spectra motivated by single-field models of inflation that would produce extra power on small scales. We find that depending on the mass and abundance of primordial black holes, it is important to include this contribution from the primordial fluctuations, so as not to underestimate the 21cm signal. Evidently our predictions of detectability, which lack any modelling of foregrounds, are unrealistic, but we hope that they will motivate improved cleaning algorithms that can enable us to access this intriguing corner of PBH-motivated parameter space.
 \end{abstract}
 
\maketitle
\section{Introduction}

The fluctuations in density left over at the end of inflation are the best probe for how inflation itself happened. Since these aren't observable directly, we must rely on mapping the evolution of these overdensities and underdensities which eventually gravitationally collapsed to form the structures that we see today. Measuring the late-time matter power spectrum will enable us to track back and predict how the fluctuations were distributed immediately after inflation, which is quantified with the primordial power spectrum. 

In order to capture the matter distribution before it is complicated by the astrophysical processes involved in reionization and galaxy formation, it is best to look at redshifts above $\sim30$. Above redshift 30, the matter in the Universe was predominantly made up of neutral hydrogen and was therefore totally dark. However, due to neutral hydrogen's spin-flip transition, the distribution of hydrogen can be detected with 21cm observations. After recombination, when the photons decoupled from the newly formed neutral hydrogen and began free-streaming towards us as the Cosmic Microwave Background (CMB), the Universe continued to cool, but Compton scattering maintained the temperature of the CMB and the gas in equilibrium. By around $z\sim200$, the Universe had cooled enough such that Compton scattering was no longer efficient enough to keep the gas and the CMB in equilibrium, and so the gas began to cool faster than the CMB. This meant that most of the neutral hydrogen was in its unexcited state, and therefore able to absorb CMB photons at the characteristic wavelength of 21cm. It is this difference in temperature of the CMB that is observable. The absorption line of the photons is redshifted from the initial wavelength of 21cm, and therefore the frequency of the radiation that arrives at detectors determines the redshift slice from which the signal originated.

\begin{figure}
    \centering
    \includegraphics[width=0.9\linewidth]{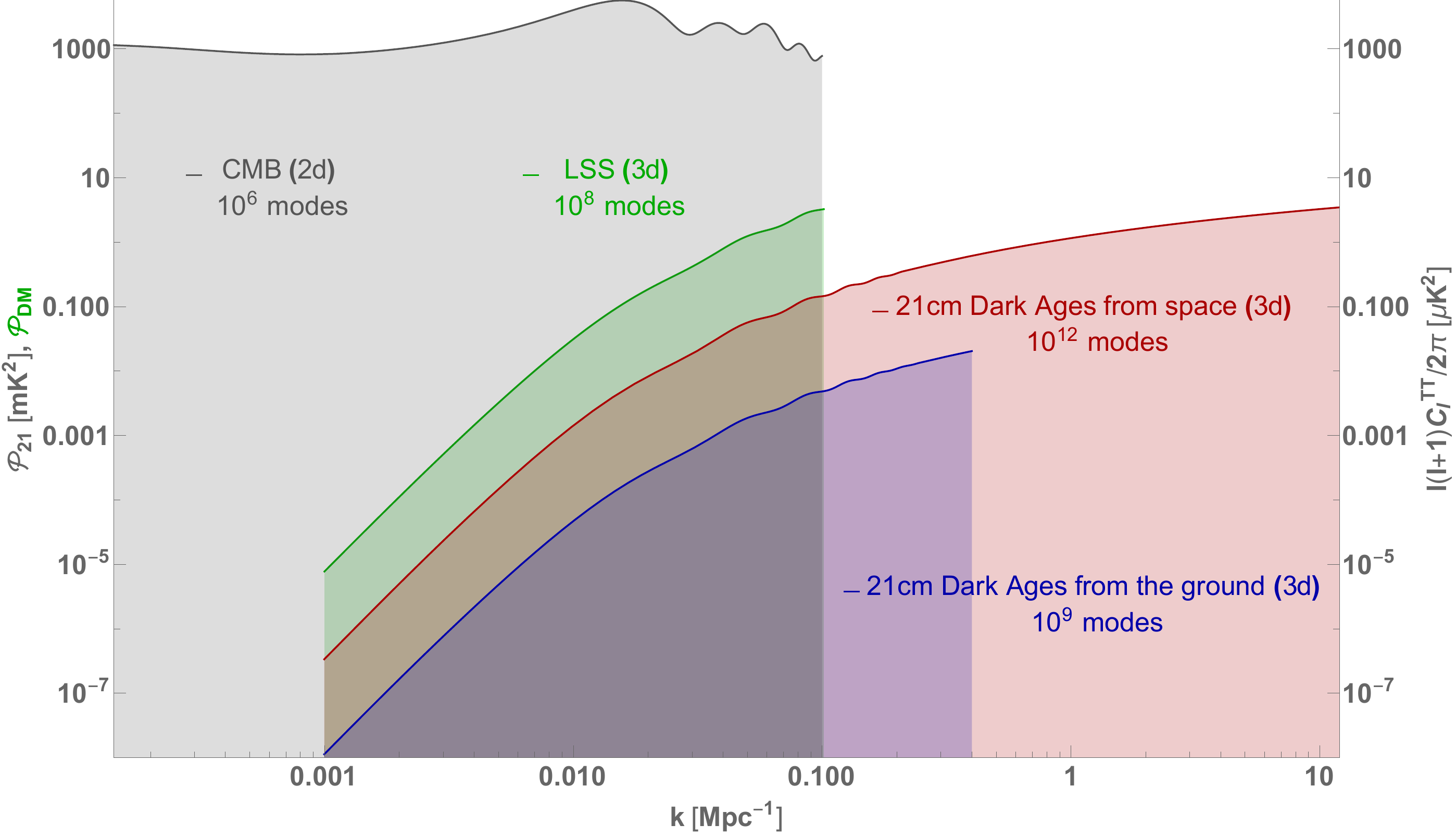}
    \caption{An illustration of the scope of different cosmological probes for accessing large numbers of modes. Note that the y-axes are different for each probe as described here. In grey is the TT angular power spectrum in units of $\mu{K^2}$ as shown on the right-hand axis, with multipoles roughly mapped to wavenumbers by $l\sim14000k/{\rm Mpc^{-1}}$ \cite{Liddle:2006ev}. In green is the dimensionless 3d matter power spectrum $\mathcal{P}_{\rm DM}$ computed with CAMB at redshift 1, which large-scale structure probes such as LSST and EUCLID will be sensitive to on scales between $k\sim0.001-0.1\,{\rm Mpc^{-1}}$ \cite{Zhan:2005rz} up to around redshift 2.5. In blue is the 3d 21cm power spectrum $\mathcal{P}_{21}$ in units of ${\rm mK^2}$ at redshift 27, which is the highest redshift accessible from ground-based experiments such as HERA and SKA. In red is the 3d 21cm power spectrum in units of ${\rm mK^2}$ at redshift 50, which would be accessible from the Moon. Note that the maximum $k$ for 21cm experiments is solely based on the angular resolution for maximum baselines given in table \ref{tab:obs}. }
    \label{fig:lss}
\end{figure}

Current CMB measurements \cite{Akrami:2018odb} constrain the primordial power spectrum very tightly on scales $k\sim10^{-4}-0.1\,{\rm Mpc^{-1}}$ to be of amplitude $2\times10^{-9}$. This means that, while a detection of the 21cm signal from the Dark Ages on any scale would be a huge achievement, it is unlikely that anything new will be uncovered about the primordial power spectrum unless smaller scales are probed. This should in theory be possible with 21cm observations if high enough redshifts can be targeted. For the best hope of a detection of the Dark Ages 21cm power spectrum, an interferometer on the Moon \cite{1985lbsa.conf..293B,Furlanetto:2019jso,Burns:2019zia} (or beyond \cite{2018EGUGA..20.3648C,Koopmans:2019wbn}) would be required to reach the small scales, that remain linear, at and above redshift 50. The constraining power of 3d 21cm power spectra measurements are illustrated in figure \ref{fig:lss}. Compared to 2d CMB measurements on large scales, and 3d large-scale structure probes on intermediate scales, both ground and space-based 21cm interferometers have the potential to access an unprecedented number of modes.

If Planck's measurements of the primordial power spectrum on large scales extrapolate to smaller scales, then the current most-favoured inflationary models (single-field, slow-roll) will continue to be preferred. However, any deviation from the low-amplitude, scale-invariant primordial power spectrum on small scales will point towards a different inflationary scenario, as well as lead to other potential observables \cite{PhysRevD.96.063503}.
For example, an enhancement in small-scale power could lead to the production of primordial black holes or ultra-compact mini haloes, which could in turn provide the seeds for supermassive black holes and the most massive galaxies \cite{Carr:2018rid,Bernal:2017nec,Latif:2016qau}. 

21cm observations can therefore teach us about both inflation and current observables at the same time. This could be complemented by a measurement of the integrated small-scale power via spectral distortions of the CMB, or by the detection of second order gravitational waves which would imply large primordial scalar perturbations, or by the detection of primordial non-gaussianity.

This paper is laid out as follows. In section \ref{sec:basics} we outline the basics of 21cm Cosmology that will enable us to produce the 21cm power spectra in section \ref{sec:power} from inflation-motivated primordial power spectra. In section \ref{sec:PBH} we demonstrate the play-off between density fluctuations produced during inflation and Poisson fluctuations in the 21cm power spectrum for different masses and abundances of PBHs and comment on their relevance with respect to accretion effects. Finally we discuss possibilities for detection in section \ref{sec:detection} and then conclude.

\section{21cm basics}\label{sec:basics}

The spin temperature $T_s$ of neutral hydrogen is defined as
\begin{equation}\label{eq:Ts}
\frac{n_1}{n_0}=3e^{-\frac{T_*}{T_s}}
\end{equation}
where $n_1$ and $n_0$ are the number densities of neutral hydrogen in excited and ground states respectively with $n_H=n_0+n_1$, and $T_*=0.068{\rm{K}}$ is the temperature corresponding to the energy difference between the ground and excited states.

In order to see how the spin temperature evolves in time we can write down the rate equations for the hydrogen atoms
\begin{equation}\label{eq:rate}
    n_0(n_H\kappa_{01}+B_{01}u_\nu)=n_1(A_{10}+B_{10}u_\nu+n_H\kappa_{10})
\end{equation}
where $A_{10}$ is the probability of spontaneous emission known as the Einstein A coefficient, $B_{10}$ is the probability of stimulated emission (when an incoming CMB photon causes another photon to be emitted and for the atom to drop from its excited to its ground state), and $B_{01}$ is the probability of stimulated absorption (when the atom absorbs a CMB photon and it jumps from its ground to its excited state). The blackbody CMB photons which mediate this process are described by the radiation field $u_\nu$. $\kappa_{10}$ and $\kappa_{01}$ are the collisional rate coefficients for which we use the values tabulated in \cite{2005ApJ...622.1356Z} - these describe the rate at which the atoms change states when they collide. In the limit of $T_*\ll{T_{\rm CMB}},{T_{s}}$, \eqref{eq:rate} can be solved and the spin temperature can be written in terms of the gas temperature, the CMB temperature, the collisional rate coefficients and the Einstein A coefficient:
\begin{equation}
    T_s=T_{\rm CMB}+(T_{\rm gas}-T_{\rm CMB})\frac{C_{10}}{C_{10}+A_{10}\frac{T{\rm gas}}{T_*}}
\end{equation}
\begin{wrapfigure}{o}{0.5\columnwidth}
    \centering
    \includegraphics[width=0.5\textwidth]{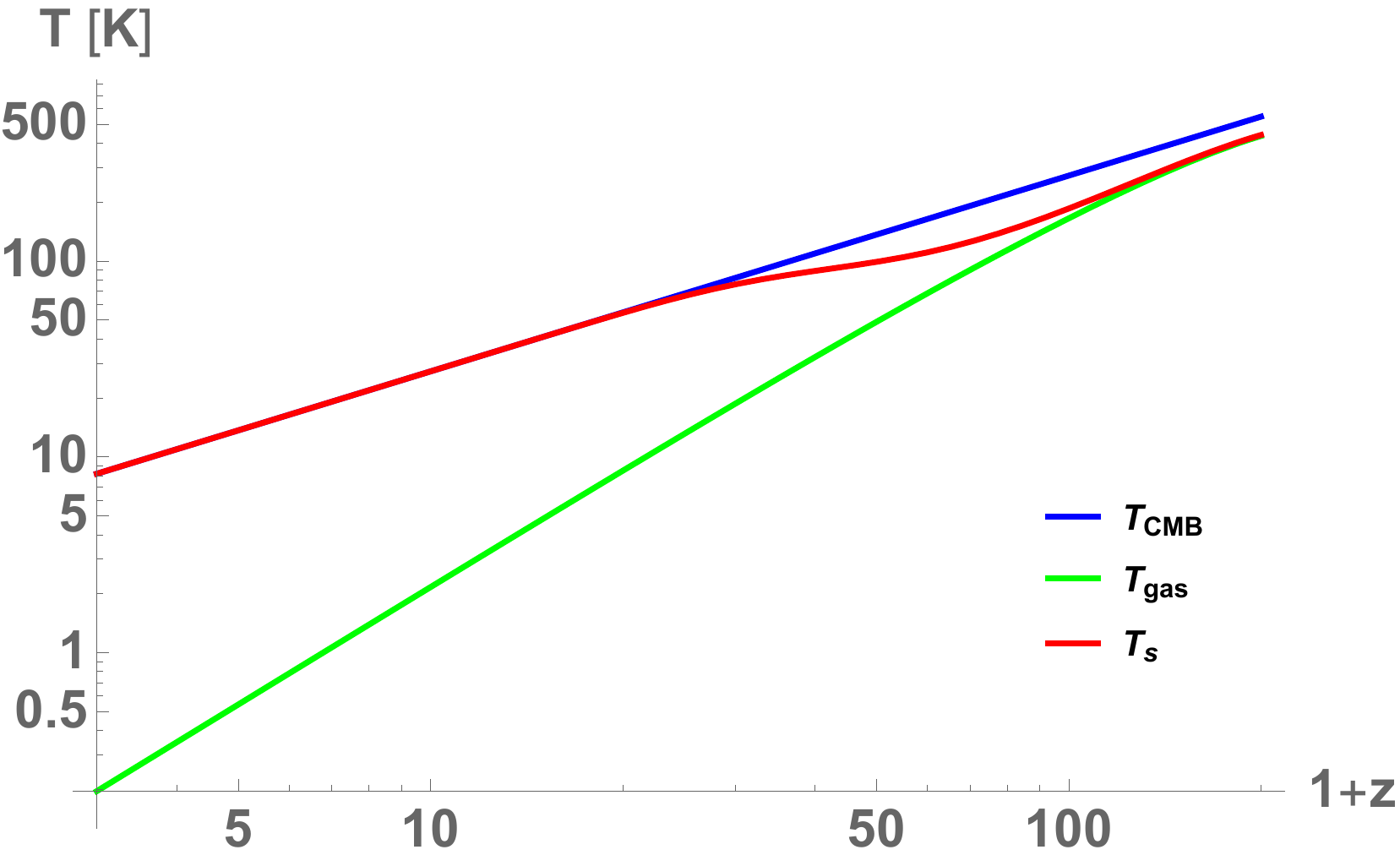}
    \caption{The evolution of the CMB temperature, gas temperature and spin temperature as a function of redshift.}
    \label{fig:Ts}
\end{wrapfigure} 
with $C_{10}=n_H\kappa_{10}$.
Figure \ref{fig:Ts} shows the evolution of the gas temperature, CMB temperature and spin temperature as a function of redshift. All three are in equilibrium until a redshift of $z\sim137$ due to residual free electrons Thomson scattering off the gas and the CMB photons. The gas then begins to cool as $T_{\rm gas}\propto(1+z)^2$ while the CMB cools as $T_{\rm CMB}\propto(1+z)$ and the spin temperature therefore deviates from both. By around $z\sim30$, the collision rate becomes subdominant to the Hubble expansion and the spin temperature couples to the CMB temperature once more. This redshift window $z\sim30-200$ is therefore the window where a 21cm signal could be observable, in absorption relative to the CMB. The quantities $T_{\rm gas}$, $n_H$, $\overline{x}$ and $T_{\rm CMB}$ are computed using RECFAST \cite{1999ApJ...523L...1S}.

The observable is not the spin temperature, but the brightness temperature $T_{21}$ which describes the contrast between the spin temperature and the CMB

\begin{equation}\label{eq:Tb}
    T_{21}=\tau\frac{T_s-T_{\rm CMB}}{z+1}
\end{equation}
where the optical depth $\tau\ll1$ depends on the neutral hydrogen density local to the absorption
\begin{equation}
    \tau=\frac{3c\lambda^2{h}A_{10}n_H}{32{\pi}k_bT_sH(z)}
\end{equation}
which can be approximated as \cite{Pillepich:2006fj}
\begin{equation}
    \tau_a=0.025\frac{T_{\rm CMB}}{T_s}\left(\frac{1+z}{51}\right)^\frac{1}{2}\left(\frac{\Omega_m}{0.27}\right)^{-\frac{1}{2}}\left(\frac{\Omega_bh}{0.035}\right).
\end{equation}
The sky-averaged brightness temperature can shed light on Cosmic Dawn and the Epoch of Reionization around redshift 10, but for the purposes of probing the scale-dependence of the 21cm signal at different redshift slices (and hence the primordial power spectrum), we will be interested in the 21cm fluctuations which track the matter fluctuations.

We will compute the 3d isotropic 21cm monopole transfer functions numerically using CAMB \cite{Lewis:1999bs}, which includes fluctuations in the density of the baryons, gas temperature, ionization fraction, radial peculiar velocities and Lyman-alpha pumping efficiency. However, fluctuations in the baryons will be largely dominant during the Dark Ages, before luminous sources have formed. The linear Boltzmann equations used in CAMB to calculate the 21cm monopole transfer functions are laid out in \cite{Lewis_2007}. The non-linear effect of the relative velocity between dark matter and baryons is not captured by CAMB. This would enhance the 21cm power spectrum on large scales $k<1\,{\rm Mpc^{-1}}$, suppress it on small scales $k>200\,{\rm Mpc^{-1}}$ and enhance it again on very large scales $k>2000\,{\rm Mpc^{-1}}$ by order unity \cite{Ali-Haimoud:2013hpa}. Since we are interested in boosts in power beyond $k>1\,{\rm Mpc^{-1}}$, and do not expect to be sensitive to scales smaller than $k\sim\mathcal{O}(10)\,{\rm Mpc^{-1}}$ even with futuristic radio interferometers, we do not include their effects here.

\section{Predictions for 21cm power spectra given different primordial models}\label{sec:power}
If the measurement of the primordial power spectrum on large scales by Planck extrapolates to small scales, it will be of the form
\begin{equation}\label{eq:CMB}
    \mathcal{P}_\mathcal{R}=A_s\left(\frac{k}{k_p}\right)^{n_s-1}
\end{equation}
with $k_p=0.05\,{\rm Mpc^{-1}}$ and $n_s\approx0.965$ \cite{Akrami:2018odb}. However, there may be an increase in power on small scales, which is theoretically motivated by the potential need to explain the seeds of supermassive black holes and the most massive galaxies, as well as the possible existence of primordial black holes or ultra compact minihaloes. 

\begin{figure}[h!]
    \centering
    \includegraphics[width=0.72\textwidth]{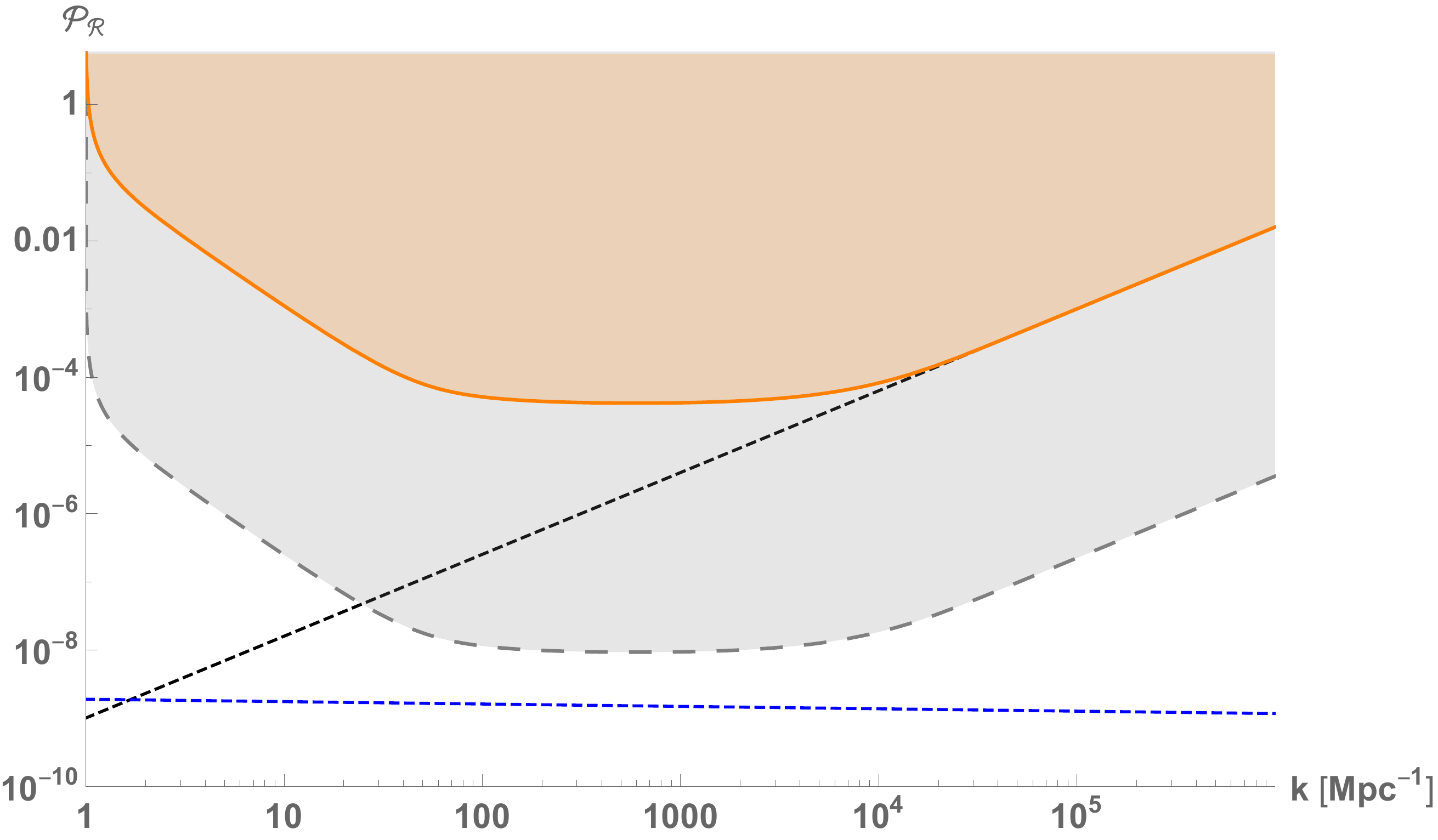}
    \caption{Constraints from COBE/FIRAS \cite{Fixsen:1996nj,1994ApJ...420..439M} on the primordial power spectrum due to $\mu$-distortions in orange - the shaded region is disallowed. Future constraints from a PIXIE-like \cite{Chluba:2019kpb} probe in grey. Constraints are calculated with an input primordial power spectrum that grows as $k^{1.2}$ with a sharp cut-off. The value of each point on the constraint curve represents the maximum amplitude that the peak of such a primordial power spectrum can be. The black dashed line grows as $k^{1.2}$ from $k=1\,{\rm Mpc^{-1}}$ and $\mathcal{P}_\mathcal{R}=10^{-9}$, i.e. the steepest that the power spectrum can be if it starts to grow at $k=1\,{\rm Mpc^{-1}}$. The blue dashed line is the canonical CDM parameterisation of the primordial power spectrum with $A_s=2.09\times10^{-9}$ and $n_s=0.965$ \cite{Akrami:2018odb}.}
    \label{fig:12mu}
\end{figure}
There are various constraints on the primordial power spectrum which must be respected. The most relevant on the scales that 21cm observations may be able to probe are those from $\mu$-distortions \cite{Fixsen:1996nj,1994ApJ...420..439M}, which constrain scales $k\sim1-10^5\,{\rm Mpc^{-1}}$. In order to avoid these, the fastest that the power spectrum can grow from $k=1\,{\rm Mpc^{-1}}$ (where CMB constraints finish) is at a rate of $k^{1.2}$. See figure \ref{fig:12mu} where the $\mu$-distortion constraints are plotted for a power spectrum that grows like $k^{1.2}$ with a sharp cut-off after the growth. The sharp cut-off is a conservative choice \cite{Byrnes:2018txb}, but if the power spectrum can't decrease that quickly \cite{Carrilho:2019oqg} then the constraints will be tighter. For single-field models of inflation with canonical kinetic terms, the fastest that the power spectrum can grow is $k^5\log{k^2}$ \cite{Carrilho:2019oqg}. However, when limited observationally by a maximum growth of seven orders of magnitude between $\mathcal{P_\mathcal{R}}\sim10^{-9}$ and $10^{-2}$, the fastest growth can be approximated by $k^4$ \cite{Byrnes:2018txb}, which also requires less restrictions on the evolution of the slow-roll parameters. The largest scale where such a fast boost can occur whilst still avoiding $\mu$-type spectral distortion constraints is $k\sim10^3\,{\rm Mpc^{-1}}$. 21cm observations offer a complimentary probe of the primordial power spectrum on these scales to spectral distortions, because they can probe the scale-dependence, whereas spectral distortion constraints are only sensitive to the integrated contribution of power across the range of scales. If an experiment such as PIXIE \cite{Kogut:2011xw} (see \cite{Chluba:2019kpb} for a recent proposal) detected a larger signal than expected from a Planck extrapolated power spectrum, the 21cm Dark Ages signal could identify which scales are contributing to the surplus.

We now find the predicted 21cm signal for 4 different primordial power spectra at redshift 27, the largest redshift accessible from the ground, and at redshift 50 when the signal is largest and would be accessible by a future lunar array. We compute the 21cm transfer functions with CAMB \cite{Lewis:1999bs}, which we combine with the four different primordial power spectra to produce the 3d 21cm power spectra.

\begin{figure}
    \centering
    \includegraphics[width=0.72\textwidth]{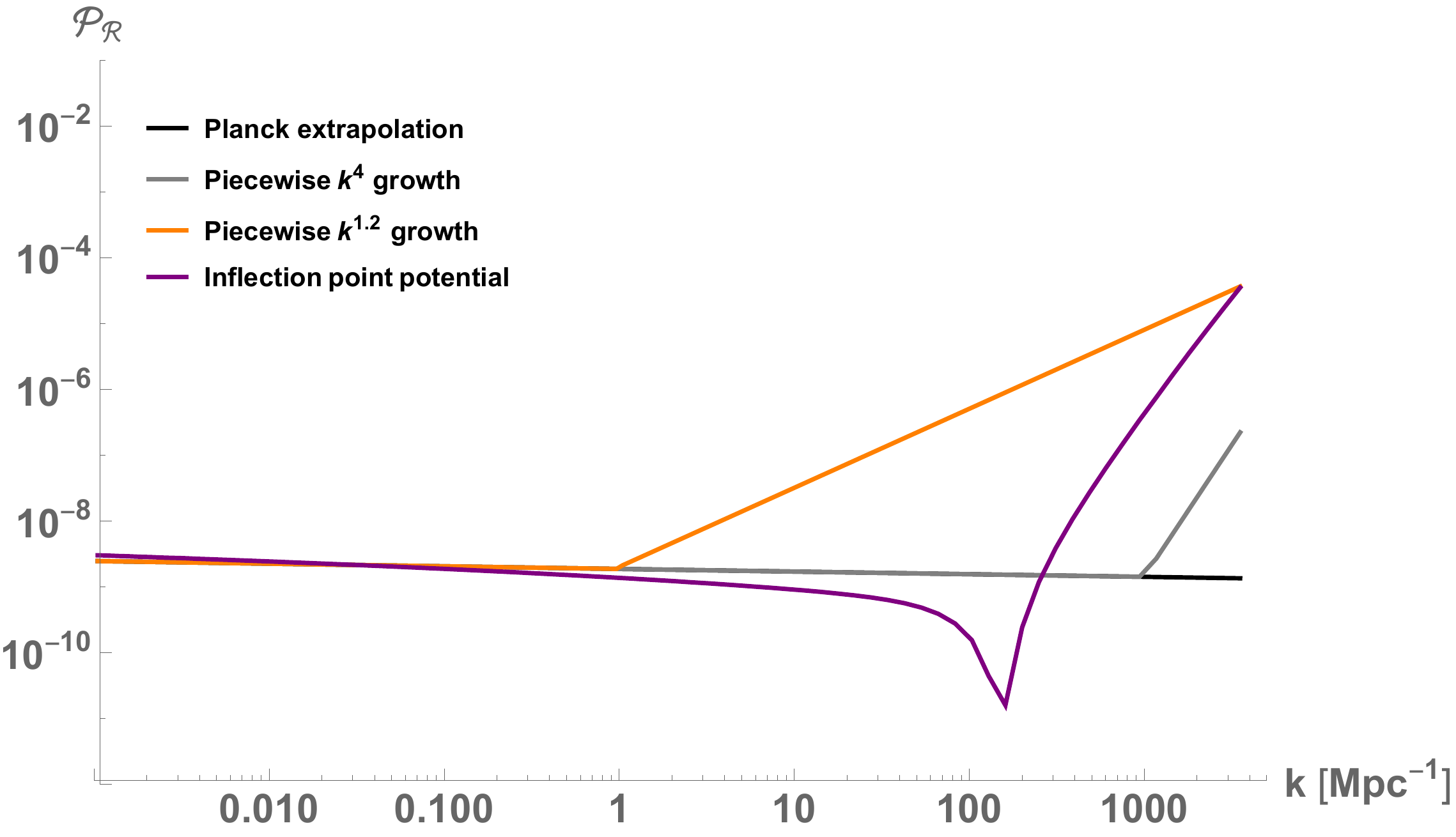}
    \caption{The primordial power spectra corresponding to the 21cm power spectra in figure \ref{kleban}.}
    \label{fig:pps}
\end{figure}

\begin{figure}
\centering
\includegraphics[width=0.72\textwidth]{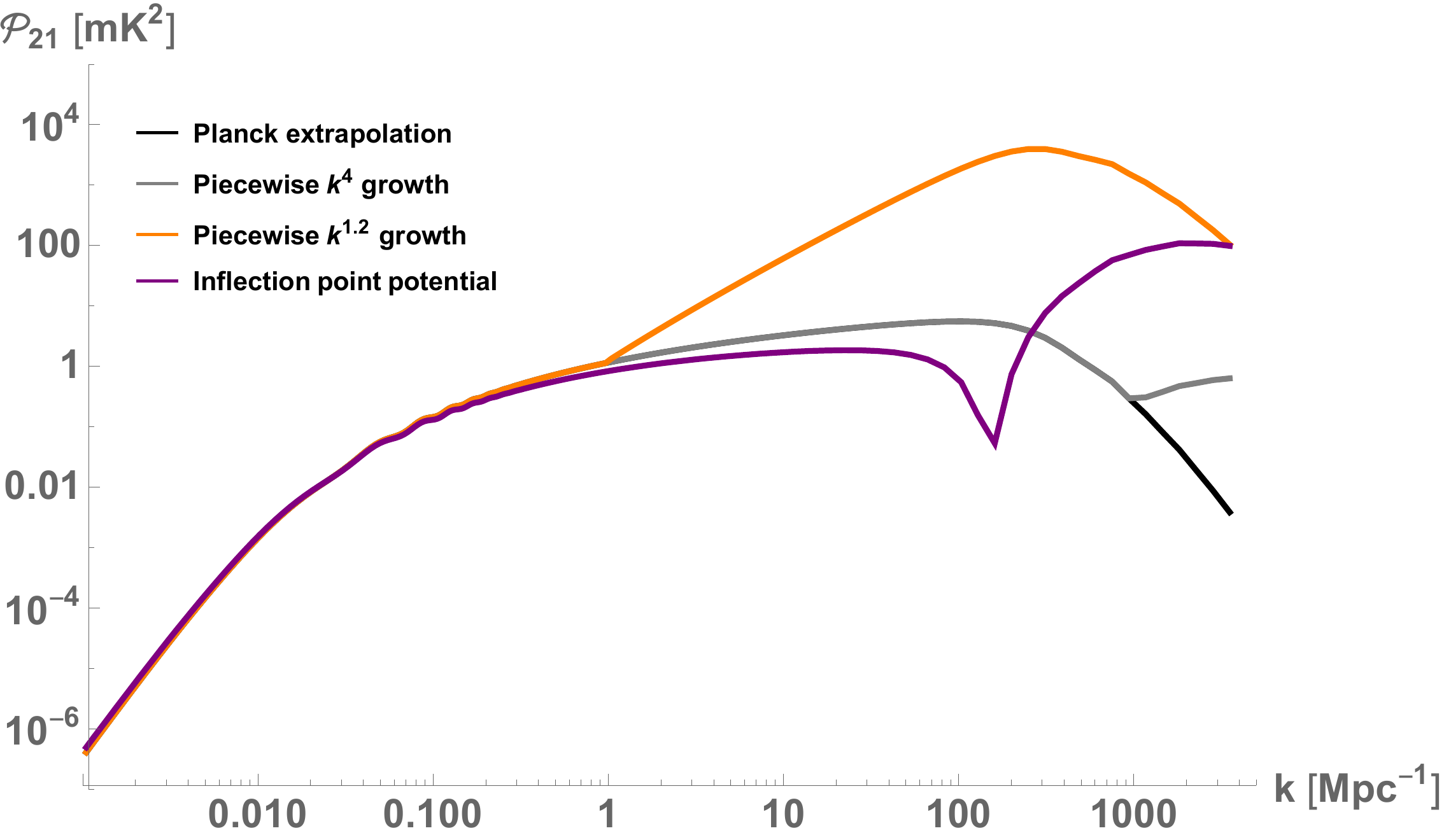}
\caption{21cm power spectrum predictions at redshift 50 for 4 different primordial power spectra as described in the text.}
\label{kleban}
\end{figure}

The four primordial power spectra chosen are shown in figure \ref{fig:pps}; in black is the spectrum extrapolated from the CMB measurements of equation (\ref{eq:CMB}), in orange is the piecewise primordial power spectrum that matches CMB measurements until $k=1\,{\rm Mpc^{-1}}$ and then grows like $k^{1.2}$ representing the maximum growth possible whilst evading spectral distortion constraints, in grey is the primordial power spectrum that matches CMB measurements until $k=1000\,{\rm Mpc^{-1}}$ and then grows like $k^4$ and in purple is the primordial power spectrum produced from a `realistic' inflationary potential \cite{Germani:2017bcs} that grows steeply on small scales before flattening off. The corresponding 21cm power spectra are plotted at redshift 50 in figure \ref{kleban}.

An excess in power can be seen for the piecewise $k^{1.2}$ growth which begins at $k=1\,{\rm Mpc^{-1}}$ because the 21cm signal has a chance to grow before it is damped at large $k$. The piecewise $k^4$ growth is just visible in comparison to the Planck-extrapolated spectrum at a scale of $k=1000\,{\rm Mpc^{-1}}$. The realistic and smooth model of \cite{Germani:2017bcs} shows a significant decrease in power which is common to inflection-point models of inflation \cite{Passaglia:2018ixg,Byrnes:2018txb}, and the subsequent growth also produces a signal in excess of the Planck model beyond $k\sim300\,{\rm Mpc^{-1}}$, although note that this could be suppressed by relative velocity effects \cite{Ali-Haimoud:2013hpa}.

If the primordial power spectrum is boosted on scales beyond $k\sim0.1\,{\rm Mpc^{-1}}$, it is plausible that 21cm interferometers will be sensitive to the signal (as well as possibly inferred from detections of the global signal \cite{Yoshiura:2019zxq,muoz2019probing}), and be able to distinguish it from the signal expected from the simplest extrapolation of the Planck measurements to  small scales. This would test whether a more complicated inflationary scenario that goes beyond the slow-roll approximation is required. We will discuss possibilities for detection in section \ref{sec:detection}.

\section{Primordial black hole production}\label{sec:PBH}
If the primordial power spectrum continues to grow on small scales beyond those plotted in figure \ref{fig:pps} until it reaches amplitudes of order $10^{-3}-10^{-2}$ \cite{Kalaja:2019uju,Young:2019yug,Young:2019osy,Kehagias:2019eil,Germani:2018jgr}, primordial black holes would be necessarily formed on that scale (corresponding to a mass via $M/M_\odot\approx(k/k_\odot)^{-2}$ \cite{Young:2019yug}) due to the collapse of large density perturbations reentering the horizon. The imprint on the 21cm signal of PBHs has been investigated before by, for example, \cite{Bernal:2017nec,Mack:2008nv,Tashiro:2012qe,Mena:2019nhm}, where the effect of accretion onto the PBHs is taken into account, as well as the Poisson fluctuations sourced by the discrete distribution of PBHs \cite{Afshordi:2003zb}. However, the primordial fluctuations that are necessary for the PBHs to form in the first place have been previously neglected. In this work, we investigate the interplay between the Poisson fluctuations and the initial fluctuations generated during inflation, as plotted in the previous section, see figure \ref{fig:pps}. We focus on regimes where accretion effects are likely to be small, and show that the primordial power spectrum cannot be neglected when calculating the 21cm signal in these cases.

Whilst PBHs will begin accreting matter at the beginning of the matter-dominated epoch, they only have an effect on the 21cm signal once the temperature of the CMB is low enough so that deviations to the spin temperature and hence the brightness temperature are noticeable. Deviations are caused by the heating and ionisation of the IGM due to the matter falling onto the PBHs. The energy is radiated by either x-ray emission or  advection-dominated accretion flow and can have both local and global effects \cite{Mena:2019nhm}. 

In this paper, our focus is on the signal at redshift 50, since it is with very low frequency radio interferometers that the smallest scales will be detectable. At redshift 50, effects of accretion on the brightness temperature are expected to be small, except for in the case of very large PBH masses and/or abundances. Since a boost in small-scale primordial power would be necessary for even just one PBH to be formed \cite{Cole:2017gle}, we consider small $f_{\rm PBH}$ so as to emphasize the importance of including the primordial power spectrum contribution, when there would be no PBH signature in the 21cm signal from accretion effects. In order to explain the seeds of supermassive black holes, only small abundances of PBHs would be required, meaning that quantifying the 21cm power spectrum in these cases is well-motivated. There may still be small (order 1) effects on the 21cm power spectrum due to accretion, however given the uncertainties in the modelling of the accretion mechanism, for example the fact that spherical accretion is assumed \cite{Mena:2019nhm}, we will not include them here. 

If instead interested in constraining PBHs as a dark matter candidate with $f_{\rm PBH}$ as close to 1 as possible, accretion effects would be imperative to understand fully and include in the calculation. In addition,  at redshifts below $z\sim30$, the effects of accretion are much more pronounced, although still heavily dependent on the mass and abundance of the PBH population. They may directly compete with the contribution from primordial fluctuations, and we leave a full investigation to future work. Given that ground-based interferometers which would be sensitive to redshifts up to $z\sim27$ cannot reach small enough scales to be sensitive to a boost in the primordial power spectrum, accretion effects may be the only way of detecting PBH signatures in the 21cm signal, as has been previously investigated \cite{Mack:2008nv,Bernal:2017nec,Mena:2019nhm,Tashiro:2012qe}.


Even if they made up all of the dark matter, the typical separation between PBHs is much larger than the comoving horizon size at the time of formation, and therefore their distribution can be described by a Poisson distribution (unlike particulate dark matter).  The Poisson fluctuations are sourced by the already-formed PBHs, and the power spectrum of the Poisson fluctuations is 
\begin{equation}
    P_{\rm Poisson}(z)=\frac{9}{4}(1+z_{\rm eq})^2D^2(z)\frac{f_{\rm PBH}^2}{n_{\rm PBH}}
\end{equation}
where $\frac{9}{4}(1+z_{\rm eq})^2$ is the transfer function for isocurvature perturbations, since they are only coupled to the dark matter content, and $z_{\rm eq}$ is the redshift of matter-radiation equality which we take to be 3449. $D(z)$ is the growth factor normalised to unity today, which we calculate using CAMB to be approximately 0.025 at redshift 50 and approximately 0.05 at redshift 27. $n_{\rm PBH}$ is the comoving number density of PBHs, and the factor $f_{\rm PBH}^2/{n_{\rm PBH}}$ can be rewritten as $f_{\rm PBH}{M_{\rm PBH}}/\Omega_{\rm DM}\rho_{\rm crit}$ which will be of importance later when we discuss the degeneracy of the mass and abundance of PBHs in the 21cm signal. The combined contribution to the matter power spectrum is then given by \cite{Afshordi:2003zb}
\begin{equation}
    \mathcal{P}_{\rm 21, combined}=\mathcal{P}_{\rm 21, adiabatic}+\frac{T_{21}^2}{T_{\rm DM}^2}\frac{k^3}{2\pi^2}P_{\rm {Poisson}},
\end{equation}
where $\mathcal{P}=k^3/2\pi^2P$ for all quantities. The combined 21cm power spectrum is then calculated by using the 21cm transfer functions, $T_{21}^2$, and the CDM transfer functions, $T_{\rm DM}$, from CAMB at a given redshift. 

\begin{figure}[h]
\centering
\includegraphics[width=0.72\linewidth]{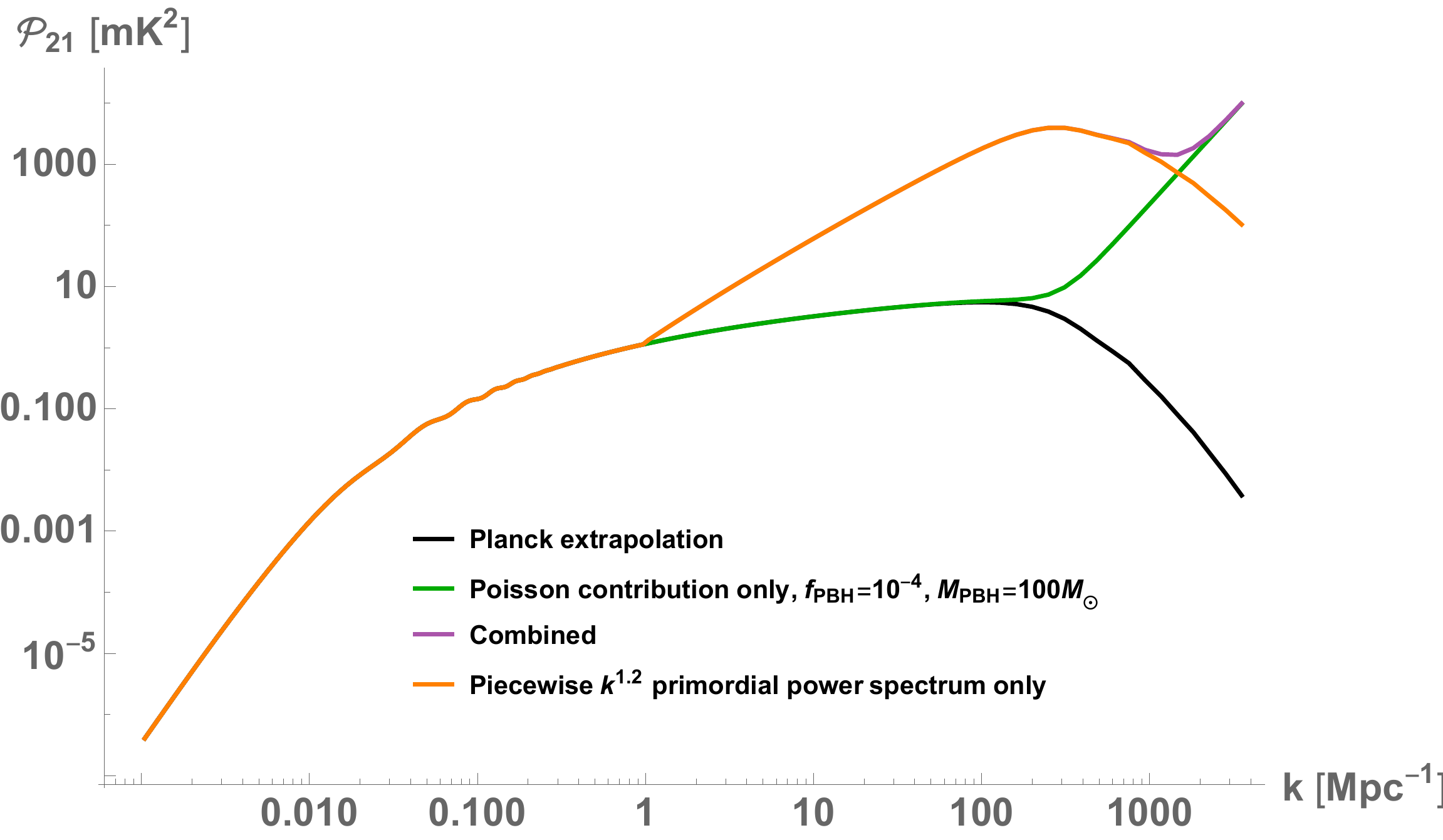}
\caption{The 21cm power spectrum at redshift 50 for the scenario where $100{\rm M_\odot}$ PBHs are produced with abundance $f_{\rm PBH}=10^{-4}$. In orange is the 21cm signal prediction taking into account just the boost in the primordial power spectrum, in green is just the Poisson contribution, and in purple is the combined result. In black is the 21cm power spectrum produced by extrapolating the primordial power spectrum measured by Planck to small scales. This demonstrates that it is important to include the primordial power spectrum boost, so as not to underestimate the 21cm power spectrum signal.}
\label{fig:poisson1}
\end{figure}

We focus on 2 different masses of PBHs, $100M_\odot$ and $10^4M_\odot$, as these are the largest and smallest mass PBHs that can be produced without conflicting with either spectral distortion or pulsar timing array constraints (e.g. \cite{Byrnes:2018txb,Inomata:2018epa}), and could be respectively produced from the primordial power spectra growing like $k^{1.2}$ and $k^{4}$ plotted in figure \ref{fig:pps}. A non-monochromatic PBH mass function is inherent in the non-monochromatic power spectra, however we assume a monochromatic mass spectrum for the PBH population in the Poisson contribution. It was shown in \cite{Byrnes:2018txb} that the mass function of PBHs produced from even very shallow primordial power spectra on the low-mass end drops off quickly. We therefore expect Poisson contributions due to extended mass functions to affect a very small range of scales larger than the peak of the PBH distribution and do not include them here. For a fuller discussion of PBHs with extended mass distributions and the 21cm signal, see \cite{Gong:2018sos}. Note that PBHs with masses $<0.1M_\odot$ can be readily produced without conflicting with any additional power spectrum constraints - their abundance is only limited by various constraints of their `direct' non-detection which vary between $f_{\rm PBH}=1$ and $f_{\rm PBH}\sim10^{-5}$ depending on the mass. PBHs with masses this low, however, would form on scales too small to  be detectable with 21cm experiments.

For $M_{\rm PBH}=100M_\odot$ the Poisson contribution is by far a sub-dominant effect for all $f_{\rm PBH}$ in comparison with the primordial power spectrum growing like $k^{1.2}$ because it occurs on much smaller scales. This means that neglecting the primordial power spectrum would predict a 21cm signal that is too small. This is shown in figure \ref{fig:poisson1}, where the 21cm power spectrum at redshift 50 is plotted. For $100M_\odot$ PBHs, the orange line is just the primordial power spectrum contribution, the green line is just the Poisson contribution, and the purple line is the combined result. On scales beyond $k\sim1\,{\rm Mpc^{-1}}$, the primordial signal is much larger than the Poisson contribution, showing that only including the Poisson fluctuations underestimates the 21cm signal if the primordial power spectrum is boosted on larger scales than the Poisson fluctuations affect. Any boost in the primordial power spectrum that occurs in the range $k\sim0.1-100\,{\rm Mpc^{-1}}$ should therefore be included in 21cm signal predictions.

\begin{figure}
    \centering
    \includegraphics[width=0.72\linewidth]{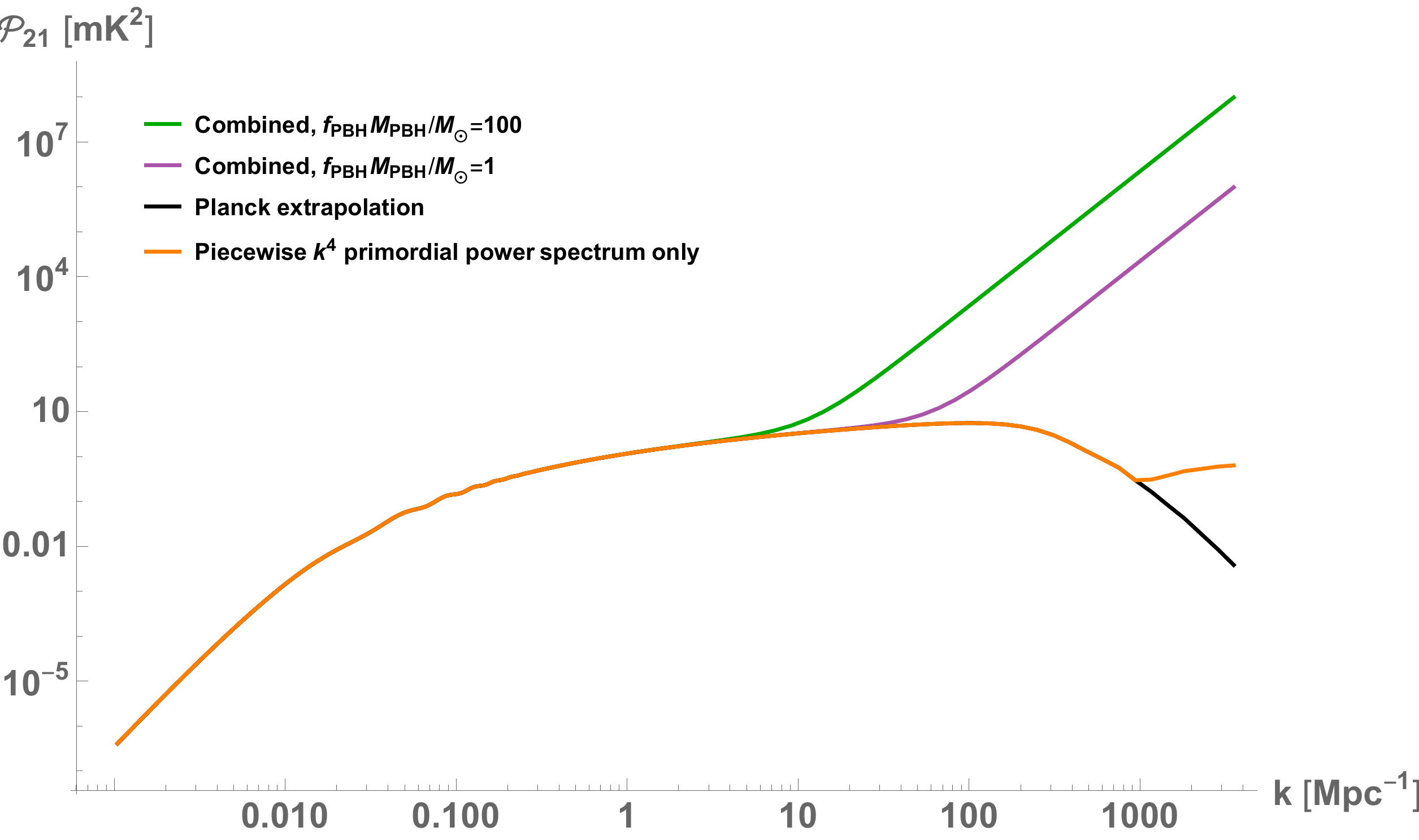}
    \caption{The 21cm power spectrum at redshift 50. The orange line only includes the primordial fluctuations contribution, for the primordial power spectrum that grows like $k^4$ and would produce $10^4M_\odot$ PBHs if extrapolated. The green and purple lines include the Poisson fluctuations for $f_{\rm PBH}M_{\rm MPBH}/M_\odot=100,1$ respectively. Since the primordial boost happens on very small scales, the Poisson contribution is dominant.}
    \label{fig:pois}
\end{figure}

The $k^4$ primordial power spectrum (grey line in figures \ref{fig:pps} and \ref{kleban}) would produce PBHs with masses around $M_{\rm PBH}=10^4M_\odot$. In this scenario, since the the primordial fluctuations grow very steeply, the boost only needs to occur on very small scales and the Poisson fluctuations generally dominate. We show this in figure \ref{fig:pois}. The orange line is just the primordial power spectrum contribution, whilst the green and purple lines show the signal including the Poisson fluctuations for the combinations $f_{\rm PBH}M_{\rm PBH}/M_\odot=100,1$ respectively. Whilst the boosted primordial fluctuations can be extrapolated to infer a most likely PBH mass produced (up to uncertainties in the mass function and horizon mass relationship), due to the degeneracy between $f_{\rm PBH}$ and $M_{\rm PBH}$ in the Poisson power spectrum, if the Poisson fluctuations dominate, the information about the mass and abundance individually is lost. In this situation, accretion effects may be able to distinguish between the two, however at redshift 50 they are likely to be small and therefore need to be accounted for very accurately. We leave an investigation of the interplay between all three effects at high redshift for future work.

\section{Possibilities for detection}\label{sec:detection}

For a rough estimate on the sensitivity of SKA to the Dark Ages 21cm signal, the crude expression from \cite{Furlanetto:2006wp} can be used which gives a scaling relation for the $1-\sigma$ error on the 21cm power spectrum at a given $k$ and $z$:

\begin{equation}\label{eq:crude}
    \sqrt{\frac{k^3\delta{P}_{21}}{2\pi^2}}\sim\frac{0.1\,{\rm mK}}{\epsilon^{1/4}f_{\rm cov}}\left(\frac{k}{0.04\,{\rm Mpc^{-1}}}\right)^{3/4}\left(\frac{T_{\rm sky}}{10^4\,K}\frac{2\,{\rm km}}{R_{\rm max}}\right)\left(\frac{10\,{\rm MHz}}{B}\right)^{1/4}\left(\frac{1000\,{\rm hr}}{t_{\rm int}}\right)^{1/2}\left(\frac{1+z}{50}\right)
\end{equation}
where $\epsilon$ is the frequency in $k$ that the data is binned, $f_{\rm cov}$ is the array covering factor, $T_{\rm sky}$ is the temperature of the
Galactic synchrotron foreground at the frequency of the observation, $R_{\rm max}$ is the radius of the (circular) array, $B$ is the bandwidth, and $t_{\rm int}$ is the number of hours of integration.

\begin{wraptable}{l}{0.5\columnwidth}
\begin{center}
\begin{tabular}{|c|c|c|c|}
     \hline
      & HERA & SKA & Lunar\\\hline$R_{\rm max}$ & 0.876km & 5km & 300km\\\hline
     $f_{\rm cov}$ &0.08 & 0.01 & 0.75\\\hline
     Bandwidth &100MHz & 2GHz& 50, 100 MHz\\\hline
     $T_{\rm sky}$ & 2000 K & 2000 K & $10^4$K\\\hline
     $t_{\rm int}$ & 1000 hours & 1000 hours &1000 hours\\\hline
     $\epsilon$ & 1 & 1 & 1\\\hline
\end{tabular}
\end{center}
\caption{\centering{Parameters describing HERA, SKA and lunar arrays.}}
\label{tab:obs}
\end{wraptable}

\begin{figure}
    \centering
    \includegraphics[width=0.72\textwidth]{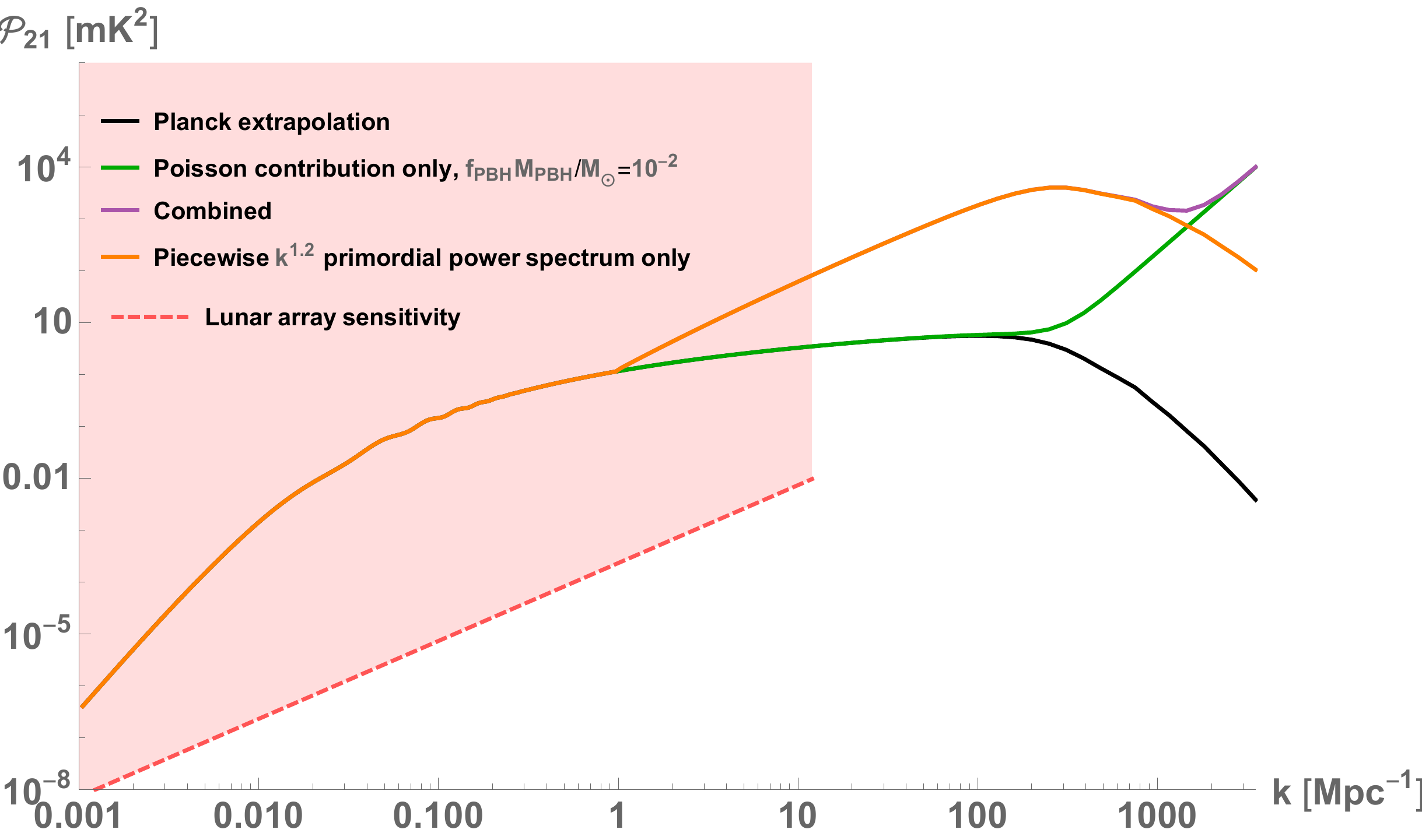}
    \caption{The same 21cm power spectrum as in figure \ref{fig:poisson1} at redshift 50. A rough estimate of the sensitivity of a possible configuration for a radio interferometer on the far side of the moon is shown by the red dashed line.}
    \label{fig:poisson2}
\end{figure}
\begin{figure}
    \centering
    \includegraphics[width=0.72\textwidth]{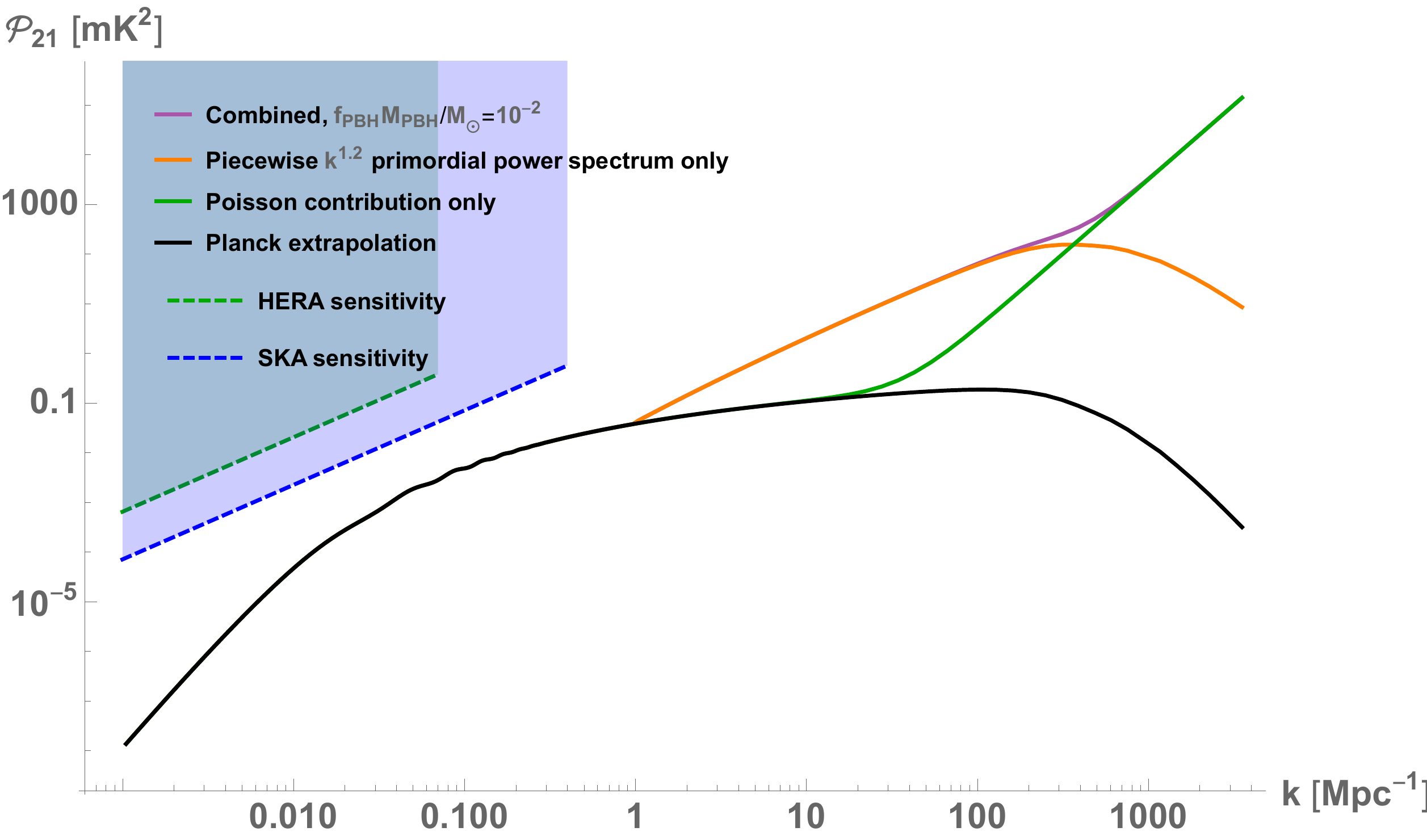}
    \caption{ The 21cm power spectrum at redshift 27 for the scenario where $100{\rm M_\odot}$ PBHs are produced with abundance $f_{\rm PBH}=10^{-4}$. In orange is the 21cm signal prediction taking into account just the boost in the primordial power spectrum, in green is just the Poisson contribution, and in purple is the combined result. In black is the 21cm power spectrum produced by extrapolating the primordial power spectrum measured by Planck to small scales. A rough estimate of the sensitivity of HERA and SKA are shown by the green and blue regions.}
    \label{fig:z27det}
\end{figure}

For the best hope of observing low frequencies, i.e. high redshifts and smaller scales, it will be necessary to go to the Moon. Using the parameters proposed by \cite{Bernal:2017nec} for a lunar radio interferometer in table \ref{tab:obs}, according to equation (\ref{eq:crude}), the sensitivity is shown by the red region in figure \ref{fig:poisson2}.

Assuming perfect foreground removal, the lunar array should be sensitive enough to measure the 21cm power spectrum at redshift 50 up to $k\sim12\,{\rm Mpc^{-1}}$. This would enable a clean distinction between the expected matter power spectrum from an extrapolation of the Planck measurements on large scales, and any deviations. Extra power, or a lack of power, on the as of yet unexplored small scales beyond $k\sim0.1\,{\rm Mpc^{-1}}$ should be observable. In addition, multiple 3d power spectra at several redshift slices could be stacked in order to increase the signal-to-noise of the detection. We demonstrate this below with a Fisher forecast for three parameters that describe a small-scale boost in power.

We have defined the smallest scale detectable as determined entirely by the angular resolution of the detector, given by $k_{\rm max}\sim2\pi{R_{\rm max}}/14000\lambda(z)\,{\rm Mpc^{-1}}$, and we have focused on the isotropic power spectrum for which the signal should be largest. However, given that foregrounds are expected to especially dominate the Fourier modes in the angular direction, $k_\perp$, (see, for example, \cite{Morales_2012}) it might be possible to reach smaller scales in the line-of-sight direction $k_\parallel$. Whilst the non-isotropic power spectrum would exhibit a smaller signal, better spectral resolution of the detector might be possible and therefore a larger $k_\parallel$ could be reached than the $k_{\rm max}$ defined by the angular resolution. A signal in the parameter space away from the foreground `wedge' would simplify the foreground removal task somewhat, however the number of independent modes lost to the wedge would decrease the signal-to-noise of any detection. See \cite{Bernal:2019jdo} for a recent investigation of using the anisotropic power spectrum to extract more cosmological information from line-intensity mapping.

Using parameters that emulate the HERA configuration and a possible SKA-Low configuration given in table \ref{tab:obs}, it is possible to put a rough estimate on the sensitivity to the 21cm signals predicted in the previous section at $z\sim27$. This is shown by the green and blue regions in figure \ref{fig:z27det}. The angular resolution means that neither HERA nor SKA-Low will be sensitive to small enough scales to go beyond the tightly constrained Planck measurements of the isotropic power spectrum up to $k\sim0.1\,{\rm Mpc^{-1}}$. This means that PBH signatures will only be detectable if the PBH masses and abundances are large so that accretion effects dominate \cite{Bernal:2017nec,Mena:2019nhm,Gong:2018sos,Tashiro:2012qe}. Note that the scaling relation (\ref{eq:crude}) does not take into account sample variance. This suffices in our case because we are interested in the sensitivity at small scales where the noise dominates, however it would be important for an accurate SKA error estimate on large scales. Furthermore, astrophysical sources would contaminate the signal at these redshifts \cite{Cohen:2016jbh}, and would need to be taken into account for an accurate prediction.

We perform a Fisher forecast for three parameters that describe a boost in the power spectrum which could be detected in the 21cm signal. We parametrise the 21cm power spectrum as

\begin{equation}
    \mathcal{P}_{21}=T_{21}^2\left(A_s\left(\frac{k}{k_*}\right)^{n_s-1}+B_s\left(\frac{k}{k_{inc}}\right)^{n_b}\right)+\frac{T_{21}^2}{T_{\rm DM}^2}\frac{9}{4}(1+z_{\rm eq})^2D^2(z)\frac{k^3}{2\pi^2}\frac{f_{\rm PBH}M_{\rm PBH}}{\Omega_{\rm DM}\rho_c}
\end{equation}
with $T_{21}$ the 21cm monopole transfer function and $T_{\rm DM}$ the cold dark matter transfer function at a given redshift, $k_{inc}$ the scale at which the primordial power spectrum is boosted from near scale-invariance, $n_b$ is the spectral index of the boosted part of the spectrum, and $B_s=A_s\left(k_{inc}/k_*\right)^{n_s-1}$. We will use $k_{inc}$, $n_b$ and $f_{\rm PBH}M_{\rm PBH}$ as the three parameters for our Fisher forecast. The Fisher matrix for the 21cm power spectrum is defined as \cite{Liu_2016}

\begin{equation}
    F_{\alpha\beta}=\sum_{k,z}\frac{1}{\varepsilon^2(k,z)}\frac{\partial\mathcal{P}_{21}(k,z)}{\partial\theta_\alpha}\frac{\partial\mathcal{P}_{21}(k,z)}{\partial\theta_\beta}
\end{equation}
with $\theta$ representing the three parameters we have chosen, $\varepsilon$ is the error on the 21cm signal given in equation (\ref{eq:crude}) and the $1-\sigma$ error bars on a single parameter we calculate with $\sigma=\sqrt{F_{\alpha\alpha}^{-1}}$. We bin \begin{wraptable}{l}{0.5\columnwidth}
\begin{center}
\begin{tabular}{|c|c|c|}
     \hline
       & $k_{\rm inc}$ & $n_b$  \\\hline
     \multirow{3}{*}{$R_{\rm max}=300{\rm km}$} &$1 \pm 0.0037$ & $1.2 \pm 0.0063$   \\
     & $5 \pm 0.22$ & $1.2 \pm 0.046$   \\
     &$5 \pm 0.18$ & $2 \pm 0.044$   \\\hline
     \multirow{3}{*}{$R_{\rm max}=500{\rm km}$} &$1 \pm 0.0022$ & $1.2 \pm 0.0038$   \\
     & $5 \pm 0.13$ & $1.2 \pm 0.028$   \\
     &$5 \pm 0.11$ & $2 \pm 0.026$   \\\hline
    \centering \multirow{1.8}{*}{$R_{\rm max}=500{\rm km}$ } &$1 \pm 0.0017$ & $1.2 \pm 0.0028$   \\
    \multirow{1.8}{*}{ $f_{\rm cov}=1$} & $5 \pm 0.098$ & $1.2 \pm 0.021$   \\
     &$5 \pm 0.081$ & $2 \pm 0.020$   \\\hline
\end{tabular}
\end{center}
\caption{\centering{$1-\sigma$ errors on fiducial values of the parameters $k_{\rm inc}$ and $n_b$ for the lunar array as described in table \ref{tab:obs}. }}
\label{table:fish}
\end{wraptable} the 21cm transfer functions in increments of $\Delta{k}=k$ to be consistent with $\epsilon=1$ in equation (\ref{eq:crude}), and we sum over three redshift slices at $z=49,50,51$ to be roughly consistent with a bandwidth of $50{\rm MHz}$ for the lunar array. We assume these slices are independent. As HERA and SKA are only likely to be sensitive to the 21cm power spectrum up to $k\sim0.07\,{\rm Mpc^{-1}}$ and $k\sim0.4\,{\rm Mpc^{-1}}$ respectively, the lunar array is the only experiment that would be sensitive to $k_{\rm inc}\geq1$. We find that the proposed specifications for the lunar array will be very sensitive to $k_{\rm inc}$ and $n_b$ but will not be able to constrain the parameter $f_{\rm PBH}M_{\rm PBH}$ at all, given that it becomes important at much smaller scales. We therefore just report the resulting $1-\sigma$ error bars for fiducial values of $k_{\rm inc}$ and $n_b$ in table \ref{table:fish}, and show the effect of varying $R_{\rm max}$ and $f_{\rm cov}$.


\section{Conclusions}
Dark ages exploration has unexcelled reach in probing excess power in the primordial spectrum on scales far smaller than those probed by the CMB or LSS, see the illustration in figure \ref{fig:lss}. Not only is this range of parameter space uniquely accessibly via 21cm spectroscopy at high $z\sim 30-80$, without any contamination from the first stars, but the huge number of modes available, further boosted by  21cm tomography,  in exploiting  power down to $k> 10 \rm \ Mpc ^{-1}$ makes this potentially the most sensitive cosmological probe possible. Of course this is a futuristic view as the foregrounds are many orders of magnitude larger at such low frequencies, ideally $\sim 30 $ MHz, amounting to a brightness temperature thousands of times larger than the elusive signal at the tens of mK level. However, CMB primordial B-mode detection faces a comparable foreground challenge, where the current CMB-S4 goal of B-mode sensitivity at the few mK level may not be insurmountable. We hence consider that it is worthwhile to develop predictions in this paper without entering into the details of the foreground limitations.

Identification of the nature of dark matter remains the highest priority in particle astrophysics and cosmology. Here we consider the principal weakly interacting candidate for non-baryonic dark matter that relies only on known physics, the primordial black hole. The challenge is to develop initial conditions in the post-inflationary universe that can produce PBHs in the empirically allowed mass range.

In order for all of the non-baryonic dark matter to be PBHs, the mass window is limited to the sub-lunar range, bounded by Hawking evaporation limits from the diffuse gamma ray background at the lower end and gravitational microlensing of M31 at the upper end,  more specifically to the mass range $10^{-17}$ - $10^{-9}\,{\rm M_\odot}.$

However, a population of larger PBHs could still be astrophysically significant even with small abundances due to the fact that they could account for the population of seed black holes required to account for the presence of supermassive black holes at $z \gtsim 6,$ namely $f_{\rm PBH}\sim 10^{-4}$. Furthermore they could account for some or all of the LIGO detections of unexpectedly massive black holes, possible if the PBH mass fraction satisfies $f_{\rm PBH}\sim 0.01$. In addition, standard PBH production scenarios require a deviance from scale-invariance in the primordial power spectrum, and therefore a detection of small-scale power would also be informative for understanding inflationary dynamics. 

We have found that PBH production in the observationally motivated range of $10-10^4 \rm M_\odot,$ requires the power spectrum to be sufficiently boosted by primordial fluctuations. If this boost occurs on larger scales, $k\sim0.1 -100{\rm Mpc^{-1}}$, this contribution must be accounted for in the 21cm power spectrum so as not to underestimate the signal. Depending on the mass and abundance, Poisson fluctuations can also become important, and in that case accurate modelling of accretion effects at high redshifts will be vital to identify the underlying PBH population producing the signal. These signatures could become potentially observable in the 21cm power spectrum with the new generation of filled low frequency interferometers. Evidently our predictions, which lack any modelling of foregrounds, are unrealistic, but we hope that they will motivate improved cleaning algorithms that can enable us to access  this intriguing corner of PBH-motivated parameter space.

\section{Acknowledgements}
PC acknowledges support from the UK Science and Technology Facilities Council via Research Training Grant ST/N504452/1. We are grateful to Nicola Bellomo, Christian Byrnes, Joshua Dillon, Guilio Fabbian, Anastasia Fialkov, Danny Jacobs, Antony Lewis, Jose Luis Bernal, Julian Munoz and Pablo Villanueva-Domingo for valuable discussions over the course of this investigation.
\bibliography{21}

\end{document}